\documentclass{aa} 
\usepackage[mathlines]{lineno}
\usepackage{blindtext}

\usepackage{graphicx}
\usepackage[varg]{txfonts}  

\usepackage{amsmath, amssymb, amsfonts}

\newcommand{\vph}{\ensuremath{v_{\mathrm{ph}}}}
\newcommand{\lacut}{\ensuremath{(L/a)_{\mathrm{cutoff}}}}
\newcommand{\lacutstep}{\ensuremath{(L/a)_{\mathrm{cutoff}}^{\rm step}}}
\newcommand{\kcstep}{\ensuremath{k_c^{\mathrm{step}}}}
\newcommand{\vainf}{\ensuremath{v_{\mathrm{A\infty}}}}
\newcommand{\vain}{\ensuremath{v_{\mathrm{A0}}}}
\newcommand{\rhoinf}{\ensuremath{\rho_\infty}}
\newcommand{\Pmax}{\ensuremath{P_{\mathrm{max}}}}
\newcommand{\Pstep}{\ensuremath{P^{\mathrm{step}}}}

\newcommand{\velunits}{~$\rm km~s^{-1}$}

\usepackage{color}

\begin{document}

\title{Standing sausage modes in coronal loops with plasma flow}
\author{Bo Li\inst{1}
	\and
	Shao-Xia Chen\inst{1}
	\and
	Li-Dong Xia\inst{1}
	\and
	Hui Yu\inst{1}
	}


\institute{Shandong Provincial Key Laboratory of Optical Astronomy and
	  Solar-Terrestrial Environment, School of Space Science and Physics,
	  Shandong University at Weihai, Weihai 264209, China
      \email{bbl@sdu.edu.cn}}

\date{} 
\abstract
{Magnetohydrodynamic (MHD) waves are important for diagnosing the physical parameters of coronal plasmas. 
Field-aligned flows appear frequently in coronal loops.}
{We examine the effects of transverse density and plasma flow structuring on standing sausage modes trapped in coronal loops, 
    and examine their observational implications in the context of coronal seismology.}
{We model coronal loops as straight cold cylinders with plasma flow embedded in a static corona. 
An eigen-value problem governing propagating sausage waves is formulated, 
    its solutions employed to construct standing modes. 
Two transverse profiles are distinguished, one being the generalized Epstein distribution (profile E) 
    and the other (N) proposed recently in~\citet{2012ApJ...761..134N}. 
A parameter study is performed on the dependence of the maximum period $\Pmax$ and cutoff length-to-radius ratio
    $\lacut$ in the trapped regime on the density parameters ($\rho_0/\rhoinf$ and profile steepness $p$)
    and the flow parameters (its magnitude $U_0$ and profile steepness $u$).}
{For either profile, introducing a flow reduces $\Pmax$ obtainable in the trapped regime relative to the static case. 
$\Pmax$ depends sensitively on $p$ for profile N but is insensitive to $p$ for profile E. 
By far the most important effect a flow introduces is to reduce the capability for loops to trap standing sausage modes:
    $\lacut$ may be substantially reduced in the case with flow relative to the static one. 
Besides, $\lacut$ is smaller for a stronger flow, and for a steeper flow profile when the flow magnitude is fixed.}
{If the density distribution can be described by profile N, then measuring the sausage mode period can help 
    deduce the density profile steepness. 
 However, this practice is not feasible if profile E better describes the density distribution.   
Furthermore, even field-aligned flows with magnitudes substantially smaller than the ambient Alfv\'en speed
    can make coronal loops considerably less likely to support trapped standing sausage modes. 
}
\titlerunning{Standing sausage modes in coronal loops}
\authorrunning{B. Li et al.}
\maketitle

\section{Introduction}
\label{sec_intro}

Capitalizing on the abundantly identified magnetohydrodynamic (MHD) waves and oscillations in the solar atmosphere,
    ``coronal seismology''~(\citeauthor{1984ApJ...279..857R}~\citeyear{1984ApJ...279..857R},
    also~\citeauthor{1975IGAFS..37....3Z}~\citeyear{1975IGAFS..37....3Z},
    \citeauthor{1970PASJ...22..341U}~\citeyear{1970PASJ...22..341U})
    proves powerful for diagnosing the atmospheric parameters that are difficult to directly yield
    (for recent reviews,
    see~\citeauthor{2012RSPTA.370.3193D}~\citeyear{2012RSPTA.370.3193D},
    and also~\citeauthor{2007SoPh..246....1B}~\citeyear{2007SoPh..246....1B},
    \citeauthor{2009SSRv..149....1N}~\citeyear{2009SSRv..149....1N},
    \citeauthor{2011SSRv..158..167E}~\citeyear{2011SSRv..158..167E} for three recent topical issues).
This rapidly growing branch of solar physics, with its applications now extended beyond
    the solar corona~\citep[e.g.,][]{2007SoPh..246....3B},
    gained its theoretical foundation with a detailed
    analysis of the modes collectively supported by a straight magnetic cylinder
    with density higher than in its surroundings
    (\citeauthor{1983SoPh...88..179E}~\citeyear{1983SoPh...88..179E}, and
    also~\citeauthor{1975IGAFS..37....3Z}~\citeyear{1975IGAFS..37....3Z}).
The axially symmetric mode, for which the azimuthal wavenumber is zero, is among the infinitely many collective modes.
When oscillating in this mode,
    a coronal loop experiences periodical changes of the loop cross-section in anti-phase with
    the density variation, with the perturbed fluid velocity primarily
    transverse~\citep[e.g.][]{2007A&A...461.1149P, 2007SoPh..246..165P, 2012A&A...543A..12G}.

{  Possibly associated with} the quasi-periodic {  oscillations} in lightcurves
    {  related to} solar flares, sausage modes can play an important role
    in seismologically diagnosing the physical parameters in the region where
    flare energy is released~\citep[see][for a recent review]{2009SSRv..149..119N}.
Their role in such a context was first postulated by~\citet{1970A&A.....9..159R}
    to account for microwave measurements,
    but extends also to the interpretation of measurements in hard X-Ray
    and white light~\citep{1982SvAL....8..132Z}.
A wealth of potential candidates in spatially unresolved radio observations that may be associated with
    sausage oscillations exists (see Table 1 compiled in~\citeauthor{2004ApJ...600..458A}~\citeyear{2004ApJ...600..458A}),
    which is enriched with spatially resolved instances
    found with the Nobeyama RadioHeliograph (NRH) data~\citep{2003A&A...412L...7N,2005A&A...439..727M,2008A&A...487.1147I}.
{  More recently, sausage oscillations were also found by~\citet{2012ApJ...755..113S}     
    using the imaging data acquired by the Atmospheric Imaging Assembly (AIA,~\citeauthor{2012SoPh..275...17L}~\citeyear{2012SoPh..275...17L})
    on board the Solar Dynamics Observatory (SDO,~\citeauthor{2012SoPh..275....3P}~\citeyear{2012SoPh..275....3P}).
They were identified in the observations made with the Rapid Oscillations in the Solar Atmosphere (ROSA) imager
    as well~\citep{2012NatCo...3E1315M}.
    }    
Among the parameters that sausage mode observations can offer,
    the magnetic field strength in the flaring region tops the list~\citep{2003A&A...412L...7N},
    while the density contrast of the flaring loop relative to its ambient corona
    and the plasma $\beta$ can be serendipitously found when fast sausage modes occur simultaneously
    with slow modes~\citep{2011ApJ...740...90V}.
{  We note that a forward modeling approach has also been adopted to examine the observability of sausage modes
    in the optically thin radiation in general~\citep{2012A&A...543A..12G},
    and in the Extreme Ultraviolet~\citep{2013A&A...555A..74A} as well as radio passbands~\citep{2014ApJ...785...86R} in particular.
It turns out that such factors as viewing angles as well as spectral, temporal and spatial resolution 
    may all be important as far as the detectability of sausage modes is concerned. 
}

Sausage modes are well known to have two distinct regimes.
Those with axial wavenumbers larger than a cutoff $k_c$ correspond to the trapped regime
    with the oscillations confined in the loop.
The leaky regime arises when the opposite is true,
    and the sausage modes are subject to damping by radiating
    their energy into the surrounding fluid~\citep{1975IGAFS..37....3Z,2007AstL...33..706K}.
However, that a sausage mode is in the leaky regime
    does not necessarily mean that it is not observationally accessible,
    provided that the quality factor $Q=\tau/P$ is sufficiently high,
    where $\tau$ and $P$ denote the damping time and period, respectively.
In the idealized case where the parameters of the loop (denoted by the subscript $0$)
    and its surroundings (subscript $\infty$)
    are piece-wise constant,
    both $k_c$~\citep{1983SoPh...88..179E} and $Q$~\citep{2007AstL...33..706K} turn out to depend primarily on the density contrast $\rho_0/\rhoinf$
    in a low beta environment.
{\bf Regarding standing sausage modes, the wavenumber cutoff $k_c$ corresponds to a critical 
    loop length-to-radius ratio $\lacut$, which separates standing modes into trapped and leaky ones.
Modes in the former (latter) category correspond to real (complex) solutions to the relevant dispersion relations.} 
When the loop radius $a$ is fixed,
    $P$ increases with increasing loop length $L$ until $L/a$ reaches the cutoff value.
A further increase in $L$ leads the sausage oscillations into the leaky regime, where $P$ turns out to increase 
    with $L$ {  as well and shows} saturation in the thin-tube
    limit ($L/a\gg 1$).
This behavior of sausage mode periods was analytically {  shown by~\citet{1975IGAFS..37....3Z,2014ApJ...781...92V}},
    and numerically demonstrated both via analyzing the relevant dispersion diagrams~\citep{2007AstL...33..706K}
    and by solving the problem as an initial-boundary value one~\citep{2007A&A...461.1149P, 2009A&A...503..569I, 2012ApJ...761..134N}.

Several aspects of the recent study by~\citeauthor{2012ApJ...761..134N}~(\citeyear{2012ApJ...761..134N}, hereafter NHM12)
    are noteworthy.
First, {  regarding a step-function transverse density profile, the maximum period that trapped sausage modes can attain
    (denoted by $\Pstep$ and explicitly given by Eq.(\ref{eq_Pmax_step}))
    is only marginally smaller than the saturation value attained in the thin-tube limit in the leaky regime
    (Fig.3 in NHM12, also Fig.2 in~\citeauthor{2014ApJ...781...92V}~\citeyear{2014ApJ...781...92V}).
Actually, solving Eq.(40) in~\citet{2014ApJ...781...92V} appropriate for the $k=0$ limit, we found that
    the period $P$ is larger than $\Pstep$ by less than $11.3\%$ for density ratios higher than $10$.
    }
Second, the tendency for the sausage mode period to increase with loop length before reaching saturation
    holds also for the smooth density profiles considered therein,
    with the saturation period increasing with the profile steepness in a sensitive manner.
{  Consequently}, the period measurements of standing oscillations may then have the potential to diagnose
    how steep the transverse density distribution is, thereby adding yet another important item
    to the list of the atmospheric parameters that can be seismologically deduced.
{  Another consequence is that, 
    $\Pstep$ as obtained in the infinitely steep case would be a good approximate
    upper limit} for the sausage mode periods {  when $\rho_0/\rhoinf \gtrsim 10$}.
Given a typical loop radius $a \sim 10^3$~km, and an internal Alfv\'en speed $\vain\ \sim 10^3$~\velunits,
    this justifies the notion that standing sausage modes are responsible primarily for second-scale
    oscillations in flare lightcurves~\citep{2004ApJ...600..458A}.

The objective of the present study is twofold.
First, does the maximal period obtainable in the trapped regime depend on the density profile
    steepness in a monotonic way for all choices of density profiles?
To this end, in addition to the profile chosen in NHM12, we also examine another profile
    which has been in wide use in the literature.
Second, what would be the effects of a field-aligned loop flow on standing sausage modes?
This is necessary given that loop flows reaching up to $\sim 100$~\velunits\
    seem ubiquitous in the corona in general~(section 4.4 in~\citeauthor{2004psci.book.....A}~\citeyear{2004psci.book.....A},
    {  also see~\citeauthor{
	2008A&A...481L..49D}~\citeyear{	
	2008A&A...481L..49D}, \citeauthor{2012ApJ...754L...4T}\citeyear{2012ApJ...754L...4T}
	for more recent results with the Hinode EUV Imaging Spectrometer}),
    and have been found in oscillating structures in particular~\citep{2008A&A...482L...9O}.
While previous studies on the flow effects are concerned mainly with kink modes~\citep{2008A&A...488..757G,2010SoPh..267..377R},
    the present study is focused on sausage ones.
Of particular interest is the behavior of the maximal period $\Pmax$ and maximum length-to-radius ratio
    allowed in the trapped regime.

Before proceeding, a few words on the approach to be used seem necessary.
We model coronal loops as a monolithic, straight, axially uniform, magnetic cylinder embedded
    in an ambient corona, and examine linear, trapped sausage oscillations
    in the framework of zero-beta, ideal MHD.
By seeing loops as being ``monolithic'', we neglect the possible effects of them being multi-threaded or involving
    other forms of structuring.
We note that for loops structured as a bundle of concentric shells, the fine structuring effect seems marginal~\citep{2007SoPh..246..165P}.
The same can be said towards the effects {  both due} to the longitudinal variation of the background
    parameters~\citep{2009A&A...494.1119P}
    and to the finite plasma beta~\citep{2009A&A...503..569I}.
Furthermore, the loop curvature is known to couple the sausage to kink modes~\citep{2000SoPh..193..139R,2009SSRv..149..299V};
    however, its influence on {  the properties} of sausage modes remains to be assessed by
    a dedicated study.
It should also be stressed that only trapped modes are to be examined.
While this approach is acceptable to find the ``dividing line'' separating the leaky and trapped regimes,
    it does not allow us to draw a definitive conclusion
    on whether the sausage periods saturate in the long-wavelength limit for general transverse density profiles
    in the case {  with flow}.
That is why $\Pmax$, the maximum period that trapped sausage modes may attain, is used.
Nonetheless, from Fig.3 in NHM12 (see also Fig.\ref{fig_period} in the present manuscript),
    it seems reasonable to speculate that a saturation exists and the saturation period is not far
    from $\Pmax$.
With this in mind, we are also interested in knowing whether $\Pmax$ may be larger than $\Pstep$ for profiles other than adopted in NHM12.
If it is indeed the case, then one expects to broaden the range of periods of the quasi-periodic oscillations
    for which standing sausage modes may account.

This manuscript is organized as follows.
In Sect.\ref{sec_model}, we offer a brief derivation of the equations governing propagating sausage waves,
    and explain the procedure for constructing standing modes.
Section~\ref{sec_numres} presents the numerical results, paying special attention to how
    a smooth transverse profile together with a loop flow affect the characteristics of the standing modes.
Finally, a summary is given in Sect.\ref{sec_conc}.

\section{Problem Formulation and Construction of Standing Sausage Modes}
\label{sec_model}

\begin{figure*}
\begin{center}
\includegraphics[width=\hsize]{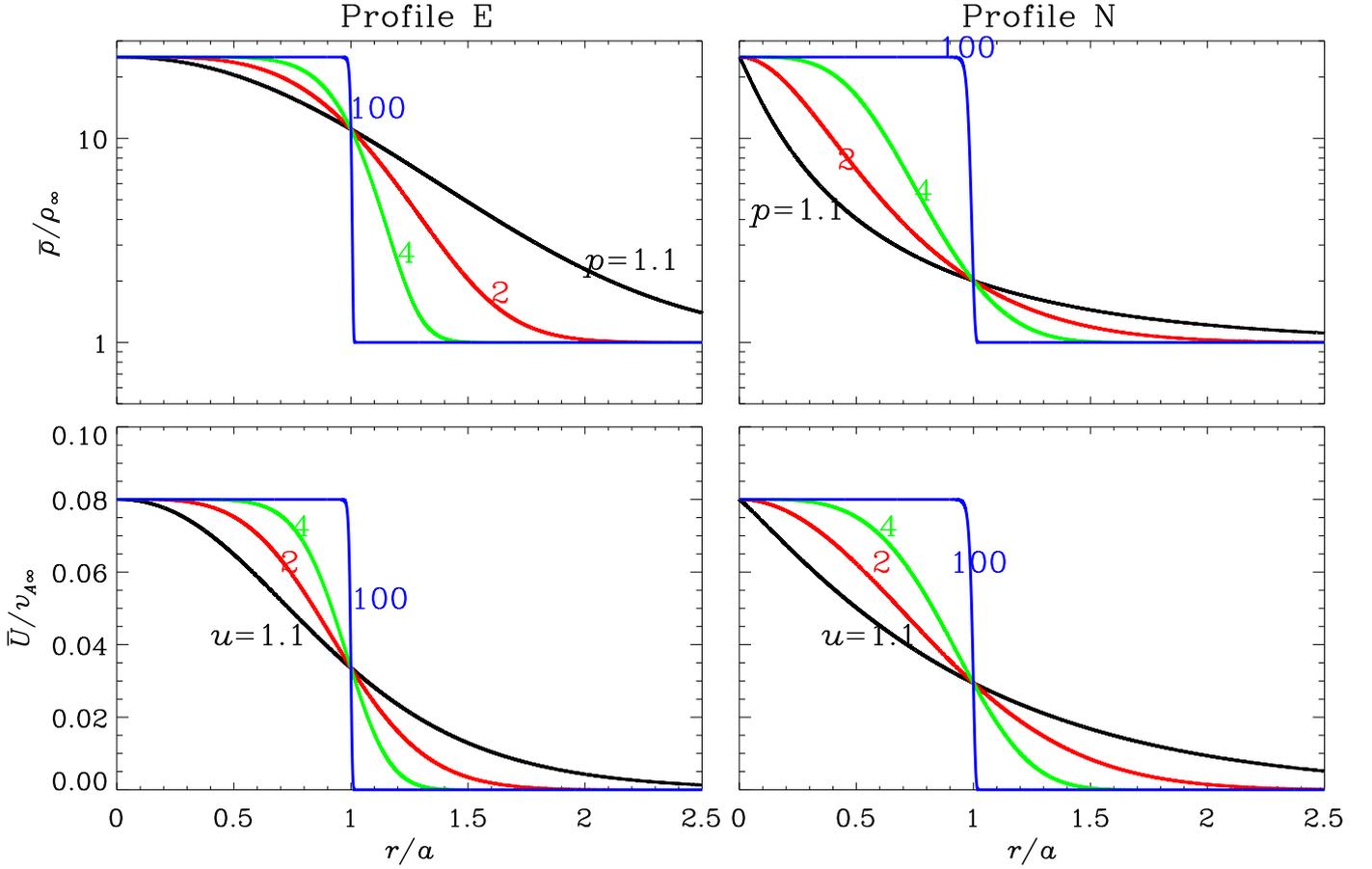}
 \caption{Background density $\bar{\rho}(r)$ and flow speed $\bar{U}(r)$ as a function of radial distance from loop axis. Two profiles are distinguished: one designated E (left column) and described by Eq.(\ref{eq_profileE}), the other designated N (right) and described by Eq.(\ref{eq_profileN}). A series of values for the density profile steepness $p$ and flow profile steepness $u$ is examined and given by different colors as denoted. For illustration, we choose a density contrast of $\rho_0/\rhoinf = 25$ and a flow magnitude of $U_0 = 0.08~\vainf$, where $\vainf$ represents the Alfv\'en speed at large distances.
}
 \label{fig_profile}
 \end{center}
\end{figure*}

The coronal loop is modeled as a straight cylinder with field-aligned flow and enhanced density embedded in
    a uniform magnetic field $\bar{\vec{B}}$.
({  The barred quantities refer to the equilibrium parameters.})    
Both the cylinder axis and $\bar{\vec{B}}$ lie
    in the $z$-direction in a standard cylindrical coordinate system $(r, \theta, z)$.
The equilibrium parameters, namely the flow speed $\bar{U}$ and background density $\bar{\rho}$,
    are structured only in the $r$-direction.
We distinguish between two profiles.
One is described by
\begin{eqnarray}
&& \bar{\rho}(r)=\rho_\infty+(\rho_0-\rho_\infty){\rm sech}^2\left[\left(\displaystyle\frac{r}{a}\right)^p\right]~, \nonumber \\
&& \bar{U}(r)   =U_0{\rm sech}^2\left[\left(\displaystyle\frac{r}{a}\right)^u\right] ,~
    \label{eq_profileE}
\end{eqnarray}
{  which is a generalized Epstein profile and called profile E for brevity.
When $p=1$, it yields the familiar symmetric Epstein profile for the density distribution, which has}
    been extensively employed in the literature
    for studying standing modes supported by static loops~\citep[e.g.][]{2003A&A...397..765C,2007A&A...461.1149P,2013SoPh..tmp..270C}.
{  The profiles} with $p >1$ were further explored by~\citet{1995SoPh..159..399N} in the context of
    impulsively excited sausage waves.
The other profile is given by
\begin{eqnarray}
    && \bar{\rho}(r)=\rho_\infty\left\{1-\left(1-\sqrt{\displaystyle
         \frac{\rho_\infty}{\rho_0}}\right)\exp
	\left[-\left(\displaystyle\frac{r}{a}\right)^p\right]
	    \right\}^{-2}, \nonumber \\
    && \bar{U}(r)=U_0\exp\left[-\left(\displaystyle\frac{r}{a}\right)^u\right]~,
	\label{eq_profileN}
\end{eqnarray}
    and is called profile N given that it was proposed in NHM12, albeit expressed in terms
    of the Alfv\'en speed profile therein.
If defining $\delta = 1-\sqrt{\rho_\infty/\rho_0}$, one readily recovers Eq.(1) in NHM12.
Both density profiles give a distribution smoothly decreasing from $\rho_0$ at $r=0$
    to $\rho_\infty$ when $r\to\infty$.
{  Consequently, the Alfv\'en speed $v_A(r) = \bar{B}(r)/\sqrt{\mu \bar{\rho}(r)}$,
    where $\mu$ is the magnetic permeability of free space,
    increases smoothly} from $v_{A0}$ at $r=0$ to $\vainf$ at large distances.
For either profile, the flow distribution is written following the same spirit as the density one.
In fact, the difference between the flow profiles is not significant,
     both yielding a distribution that decreases from $U_0$ at $r=0$
     to zero far from the cylinder, representing a loop {  with flow} placed in a static ambient corona.
In Fig.\ref{fig_profile}, the two profiles are depicted for a series of $p$ and $u$, with
     the combination of $[\rho_0/\rhoinf, U_0]$ taken to be $[25, 0.08~\vainf]$
     for illustrative purpose.
Evidently, for both profiles, the larger the steepness $p$ or $u$, the closer they are to
    a step function.

{  
For future reference, let us summarize the relevant information in the zero-beta MHD on sausage modes supported by static loops
    where the transverse density distribution is of a step-function form.
The quantities are to be denoted by a superscript ``step''.
First, the cutoff wavenumber is given by
\begin{eqnarray}
  \kcstep = \frac{j_{0,0}}{a \sqrt{\rho_0/\rhoinf -1}}~,
  \label{eq_kc_cut_step}
\end{eqnarray}
    where $j_{0,0} = 2.4048$ is the first zero of Bessel function $J_0$~\citep{1983SoPh...88..179E}.
The corresponding maximum length-to-radius ratio, only below which the loops can support trapped sausage modes, 
    is given by
\begin{eqnarray}
  \lacutstep = \frac{\pi}{\kcstep\ a} = \frac{\pi}{j_{0,0}} \sqrt{\frac{\rho_0}{\rhoinf} -1}.
  \label{eq_la_cutstep}
\end{eqnarray}
Letting $\Pstep$ denote the period attained at this $L/a$, one finds
\begin{eqnarray}
  \Pstep = \frac{2\pi}{j_{0,0}}\frac{a}{\vainf}\sqrt{\frac{\rho_0}{\rhoinf} -1} ,
  \label{eq_Pmax_step}
\end{eqnarray}
    which is well approximated by $2.62 a/v_{A0}$ for high density contrasts.
}

We adopt the ideal, zero-beta, MHD equations to describe axisymmetrical ($\partial/\partial\theta\equiv 0$)
    sausage waves.
Let $\vec{v}$ and $\vec{b}$ represent the magnetic field and velocity perturbations, respectively.
The relevant equations are then
\begin{eqnarray}
&& \frac{\partial {\vec{v}}}{\partial t}
    +\bar{\vec{U}}\cdot \nabla \vec{v}
    +\vec{v}\cdot\nabla\bar{\vec{U}}
   = \frac{1}{\mu \bar{\rho}}\left(\nabla\times \vec{b}\right) \times \bar{\vec{B}} ,
   \label{eq_momen} \\
&&  \frac{\partial \vec{b}}{\partial t}
    = \nabla\times\left(\bar{\vec{U}}\times \vec{b} + \vec{v}\times \bar{\vec{B}} \right) .
   \label{eq_Faraday}
 \end{eqnarray}
Appropriate for sausage waves, a perturbation $f(r, z; t)$ may be Fourier decomposed in $z$ and time $t$,
\begin{eqnarray}
    f(r,z;t)={\rm Re}\{\tilde{f}(r)\exp[-i(\omega t-kz)]\},
    \label{Fourier}
\end{eqnarray}
where $k$ is the axial wavenumber, and $\omega$ is the angular frequency.
The phase speed is then $v_{\mathrm{ph}} = \omega/k$.
Letting $' = \mathrm{d/d}r$,
    one finds from Eqs.(\ref{eq_momen}) and (\ref{eq_Faraday}) that
\begin{eqnarray}
     k(v_{\mathrm{ph}} -\bar{U}) \tilde{v}_r
&=& -\frac{\bar{B}}{\mu \bar{\rho}}
       \left(i \tilde{b}_z' + k \tilde{b}_r\right) ,
        \label{eq_tilde_vr} \\
  -k(v_{\mathrm{ph}} -\bar{U}) (i\tilde{b}_z)
&=& \tilde{b}_r \bar{U}' - \frac{\bar{B}}{r}\left(r\tilde{v}_r\right)' ,
        \label{eq_tilde_bz} \\
  -(v_{\mathrm{ph}} -\bar{U}) \tilde{b}_r
&=& \bar{B} \tilde{v}_r .
	\label{eq_tilde_br}
\end{eqnarray}
Note that the $z$-component of the momentum equation is not relevant, although $\tilde{v}_z$
    does not vanish when $\bar{U}'$ is not zero.
With the aid of Eq.(\ref{eq_tilde_br}), one readily sees that Eq.(\ref{eq_tilde_bz}) is equivalent to
    $i k \tilde{b}_z = -(r \tilde{b}_r)'/r$, namely the requirement that
    $\nabla\cdot\vec{b} = 0$.
Using this relation together with Eq.(\ref{eq_tilde_br}) to eliminate $\tilde{v}_r$ and $\tilde{b}_z$ from
    Eq.(\ref{eq_tilde_vr}), one finds
\begin{eqnarray}
    [\vph-\bar{U}(r)]^2\tilde{b}_r(r)=-\frac{v^2_{\rm A}(r)}
    {k^2}\left[\frac{\mathrm{d}^2}{\mathrm{d}r^2}
    +\frac{1}{r}\frac{\mathrm{d}}{\mathrm{d}r}
    -\left(k^2+\frac{1}{r^2}\right)\right]\tilde{b}_r(r) .
    \label{eq_eigen_br}
\end{eqnarray}

The boundary conditions required for formulating a standard eigen-value problem
     for sausage waves are usually specified in terms of the radial velocity perturbation $\tilde{v}_r$.
To be specific, these are $\tilde{v}_r (0) = 0$ and $\tilde{v}_r$ approaches zero sufficiently rapidly when $r$ approaches infinity.
In view of Eq.(\ref{eq_tilde_br}), they translate to
\begin{eqnarray}
  \tilde{b}_r( r = 0) = 0~, \hspace{0.2cm}
  \tilde{b}_r (r \to \infty) \to 0~.
\label{eq_eigen_BC}
\end{eqnarray}

Equation~(\ref{eq_eigen_br}) supplemented with the boundary conditions~(\ref{eq_eigen_BC})
    can be readily solved by standard numerical routines with the axial wavenumber $k$ seen as a parameter
    and $\vph$ as an eigen-value.
In practice, the code we use is a MATLAB boundary-value-problem solver BVPSUITE
    in its eigen-value mode~\citep{2009AIPC.1168...39K}.
We performed an extensive test of the code using available analytical solutions
    to known eigen-value problems 
    {  in the context of coronal seismology, and found excellent agreement between
    the numerical and analytic results}    
    for an extensive range of density parameters.
{ (For details, please see the appendix).}    
We note that the solution to Eq.(\ref{eq_eigen_br}), expressing $\vph$ as a function of $k$,
    is uniquely determined once one chooses a profile, E or N,
    for the background density and flow speed, and specifies a combination
    of dimensionless parameters $[\rho_0/\rho_\infty, p; U_0/v_{A\infty}, u]$.
For both $p$ and $u$, an extensive range of $[\sim 1, 100]$ is explored in the present work.
As for $\rho_0/\rhoinf$, a range of $[4, 100]$ is examined, covering both active region
    and flare loops.
Given that typically coronal loop flows are sub-Alfv\'enic, we consider only the values of $U_0$
    that are smaller than $0.08~\vainf$.
We further note that while Eq.(\ref{eq_eigen_br}) permits multiple solutions for large $k$,
    we always choose the one such that the corresponding eigen-function
    possesses only one extremum.
This corresponds to the branch with the smallest
    cutoff wavenumber~\citep[e.g., Fig.4 in][]{1983SoPh...88..179E}.

\begin{figure}
\begin{center}
  \includegraphics[width=\hsize]{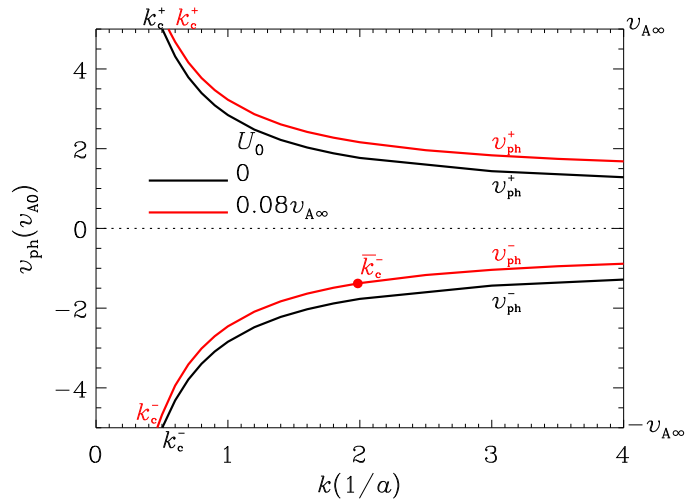}
 \caption{Axial phase speed $\vph$ as a function of longitudinal wavenumber $k$ for a loop with profile N. The black and red curves correspond to the static {  case} ($U_0=0$) and {  a case with flow} ($U_0=0.08~\vainf$ and $u=100$), respectively. Here for the density parameters, a contrast of $25$ and 
 {\bf a steepness $p=100$ are chosen}. The horizontal dotted line corresponds to $\vph = 0$. 
 {  The curves in the first (fourth) quadrant are labeled $\vph^+$ ($\vph^-$)}.
 The {  rest of the} symbols represent the wavenumber cutoffs (please see text for details).}
 \label{fig_DR}
 \end{center}
\end{figure}

Figure~\ref{fig_DR} presents, using profile N as an example, the variation of
    $v_{\rm ph}$ with $k$ for {\bf the combination of parameters
    $[\rho_0/\rho_\infty, p] = [25, 100]$}.
The black lines are for the static case ($U_0=0$),
    while the red curves are for a case {  with flow} with $U_0 = 0.08 v_{A\infty}$
    and $u=100$.
The horizontal dotted line represents $v_{\rm ph} = 0$.
It can be seen from Fig.\ref{fig_DR} that {  the curves} in the first { (labeled $\vph^+$)}
    and fourth (labeled $\vph^-$) quadrants
    are symmetric about $v_{\rm ph} = 0$ in the static case. However, the symmetry is absent in the presence of
    a background flow.
In particular, the cutoff wavenumber, only beyond which trapped sausage waves can be supported,
    is shifted towards a larger (lower) value in the first (fourth) quadrant.

Suppose we have a pair of propagating waves with axial wavenumbers $[k^+, -k^-]$
    and angular frequencies $[\omega^+, \omega^-]$,
    where both $k^+$ and $k^-$ are positive,
    and $\omega^+ = k^+ \vph^+(k^+), \omega^- = k^- |\vph^-(k^-)|$.
Furthermore, let the loop of length $L$ be a segment located between $z=0$ and $z=L$.
Let us briefly explain how to construct standing sausage modes
    (for details, see~\citeauthor{2013ApJ...767..169L}~\citeyear{2013ApJ...767..169L}).
It was shown by~\citet{2010SoPh..267..377R} that the appropriate axial boundary condition for
    standing transverse modes is that the two loop ends
    are two permanent nodes for {  the radial Lagrangian displacement at the tube boundary,
    $\xi(r=a, z; t)$}.
While the rigorous derivation therein was intended for kink modes, it also holds for standing sausage modes,
    which are also transverse in nature.
For $\xi(r=a, z=0; t)$ and $\xi(r=a, z=L; t)$ to be zero at arbitrary $t$, one naturally requires
    that $\omega^+ = \omega^- = \omega$, i.e., $k^+ \vph^+(k^+) = k^- |\vph^-(k^-)|$,
    where the phase speed for $k^{+}$ ($k^{-}$) is evaluated along the branch 
    in the first (fourth) quadrant in Fig.\ref{fig_DR}.
Consequently, $\xi(r=a, z; t)$ is in the form    
\begin{eqnarray}
  \xi(a, z; t) 
&\propto& \cos\left(\omega t - k^+ z\right) - \cos\left(\omega t +k^{-} z\right) \nonumber \\
&\propto& \sin\left(\frac{k^+ + k^-}{2}z\right)\sin\left(\omega t-\frac{k^+ - k^-}{2}z\right) .
\label{eq_stand_displc}
\end{eqnarray}
One then needs to require that
\begin{eqnarray}
k^+ + k^- =  \frac{2\pi n}{L}, n=1, 2, \cdots
\label{eq_k4stand}
\end{eqnarray}
    {\bf in order to meet the boundary conditions.}
By convention, $n=1$ corresponds to the fundamental mode, and $n \ge 2$ to its overtones.
Throughout this manuscript, we focus on the fundamental mode, which is important from an observational viewpoint since it yields
    the longest period.

\begin{figure}
\begin{center}
  \includegraphics[width=\hsize]{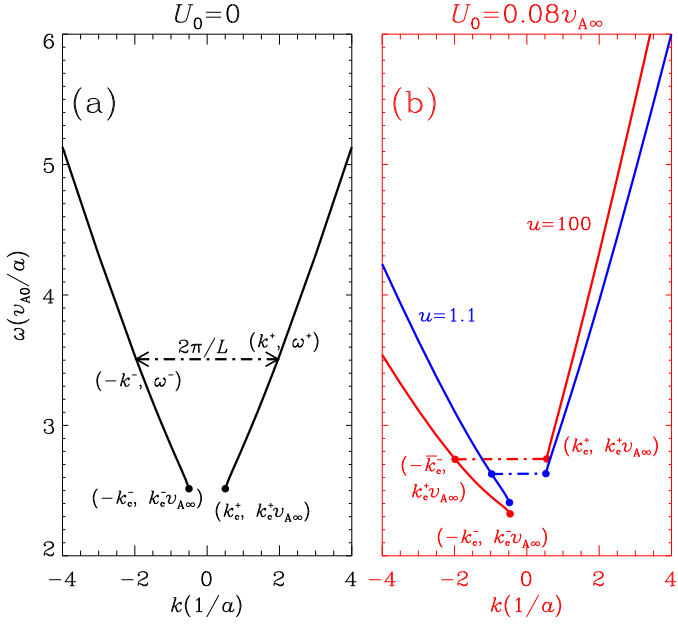}
 \caption{Dependence of angular frequency $\omega$ on axial wavenumber $k$
     for both (a) a static loop and (b) a loop with flow.
 The numerical results found in Fig.\ref{fig_DR} are adopted. 
 The $\omega-k$ curve in the first (second) quadrant derives from the curve in the 
     first (fourth) quadrant of Fig.\ref{fig_DR}.     
 {\bf In panel (b), the blue curves correspond to the case with a different flow profile steepness, $u=1.1$.}
 }
 \label{fig_stand_const}
 \end{center}
\end{figure}

Figure~\ref{fig_stand_const} illustrates how the standing modes are constructed in a simple graphical manner,
    for both (a) a static loop and (b) a loop with flow.
Here the computed results in Fig.\ref{fig_DR} are used, which are now converted into the $\omega-k$ space.
{\bf (The blue curves in Fig.\ref{fig_stand_const}b correspond to a different flow profile steepness, $u=1.1$.
They are to be discussed in relation to Fig.\ref{fig_period_p}.)}
Note that instead of mapping the curves in the fourth quadrant in Fig.\ref{fig_DR} to the fourth quadrant
    with positive $k$ but negative $\omega$,
    we map them to the second quadrant with negative $k$ but positive $\omega$.
Given a loop length $L$, in view of Eq.(\ref{eq_k4stand}) one readily finds the corresponding $k^+$ and $k^-$ by drawing a horizontal 
    line in the $\omega-k$ diagram and then measuring the separation between the two intersections of this horizontal line
    with the $\omega^+$ and $\omega^-$ curves.
The fundamental mode corresponds to the case where this separation equals $2\pi/L$
    (see the horizontal dash-dotted line in Fig.\ref{fig_stand_const}a).
Evidently, for static loops, the $\omega^+$ and $\omega^-$ curves are symmetric with respect to the $k=0$ axis,
    and consequently both $k^+$ and $k^-$ are simply $\pi/L$.
For loops with flow, for a given $L$ one has to find $k^+$ and $k^-$ numerically though, given that
    a symmetry between the $\omega^+$ and $\omega^-$ curves is no longer present.
Let $k^+_c$ and $k^-_c$ denote the cutoff wavenumbers in the first and fourth quadrants in Fig.\ref{fig_DR}, respectively.
For static loops, one finds that $\lacut$ is simply $2\pi/[(k_c^+ + k_c^-)a] = \pi/(k_c^+ a)$.
However, when $U_0$ is not zero, then one finds that $\lacut$ is not determined by $2\pi/[(k_c^+ + k_c^-)a]$
    but by $2\pi/[(k_c^+ + \bar{k}_c^-)a]$, where $\bar{k}_c^-$ satisfies
    $\bar{k}_c^- |\vph^-(\bar{k}_c^-)| = k_c^+ \vainf$ (Fig.\ref{fig_stand_const}b, the derived 
    $\bar{k}_c^-$ is also given by the filled dot in Fig.\ref{fig_DR}).
Evidently, $\lacut$ in this case {  with flow} is smaller than its static counterpart.
As for the maximum period allowed for trapped modes, it is determined by $2\pi/(k_c^+ \vainf)$ for
    the static and non-static cases alike.
To illustrate this point, let us give some specific values.
For the static case {\bf (with parameters $\rho_0/\rhoinf = 25$ and $p=100$), we find that $k_c^+ = k_c^- = 0.503/a$,
    while for the non-static case ($U_0 = 0.08\vainf$ and $u=100$), 
    $k_c^+ = 0.549/a$ and $k_c^- = 0.464/a$.
Consequently, for the non-static case, one finds that
    $\Pmax = 2\pi/(k_c^+ \vainf) = 2.291 a/\vain$,
    which is not far from the static case where $\Pmax=2.5 a/\vain$.
However, in the non-static case $\bar{k}_c^{-}= 1.994/a$, resulting in
    $\lacut = 2\pi/[(k_c^+ + \bar{k}_c^-)a] = 2.471$, which is significantly smaller than 
    in the static case where $\lacut = \pi/(k_c^+ a) = 6.246$.}
For comparison, the cutoff wavenumber in the static case with the same density ratio but a discontinuous density profile,
    $\kcstep = 0.491/a$, and consequently $\Pstep = 2.56 a/\vain$ and 
    $\lacutstep = 6.4$ (see Eqs.(\ref{eq_kc_cut_step}) to (\ref{eq_Pmax_step})).
At this point, it should be pointed out that reversing the sign of $\bar{U}$ does not change the periods and cutoff loop-to-radius ratios,
    for in that case, the $\omega-k$ diagrams will be a mirror-reflection of Fig.\ref{fig_stand_const},
    with $\Pmax$ and $\lacut$ determined by $2\pi/(k_c^- \vainf)$ and $2\pi/[(k_c^- + \bar{k}_c^+)a]$, respectively,
    where $\bar{k}_c^+$ is determined by $\bar{k}_c^{+}\vph^+(\bar{k}_c^{+}) = k_c^-\vainf$.
As detailed in the appendix in~\citet{2013ApJ...767..169L},
    this statement is true as long as the ambient corona is static.

Two things are clear from the construction procedure.
First, from Eq.(\ref{eq_stand_displc}) it immediately follows that in half of each period the radial displacement at the loop boundary
      possesses an additional node, which moves between $z=0$ and $z=L$.
In addition, the phase of the mode depends linearly on the distance along the loop.
These two signatures are not specific to sausage modes but common to standing modes in a loop with flow
    (see Fig.1 in~\citet{2011ApJ...729L..22T} where standing kink modes are of interest).
Actually, \citet{2011ApJ...729L..22T} demonstrated that these signatures are indeed in line with the kink modes observed with TRACE
    and EIT by~\citet{2010ApJ...717..458V}.
However, their detection in sausage modes has yet to be reported.
A forward modeling approach, similar to~\citet{2012A&A...543A..12G,2013A&A...555A..74A,2014ApJ...785...86R},
    would help in elucidating the detectability of these signatures.
Second, while trapped standing modes are the main concern here, one may speculate what happens when one or both propagating waves
    are leaky.
If both of them are leaky, the resulting standing mode would be leaky as well.
If one is trapped but the other leaks out by transmitting waves into the surrounding fluid,
    a standing mode will be unlikely: in the end only the trapped propagating wave would survive.
These aspects certainly merit a dedicated study, which is beyond the scope of the present manuscript though.

\section{Numerical Results}
\label{sec_numres}

We are now in a position to address how the introduction of smooth profiles, as opposed to
   step-function forms, affects the period of trapped standing sausage modes,
   and how it affects the transition line separating the trapped and leaky regimes.

\begin{figure}
\begin{center}
 \includegraphics[width=\hsize]{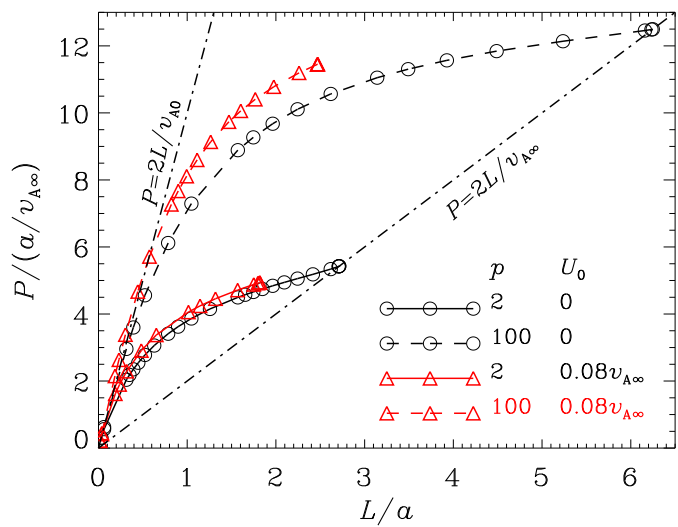}
 \caption{Sausage mode period $P$ as a function of loop length measured in units of loop radius for a loop described by profile N. The black and red curves are for the static {  case} (flow magnitude $U_0=0$) and
   {  a case with flow} ($U_0=0.08~\vainf$ with steepness $u=100$), respectively. A density contrast of $25$ is adopted. However, two values of density steepness $p$ are examined, and given by the solid ($p=2$) and dashed ($p=100$) curves, respectively. The two dash-dotted lines represent the two limiting cases for static loops where the phase speed equals either the internal or external Alfv\'en speed.}
 \label{fig_period}
 \end{center}
\end{figure}

\begin{figure}
\begin{center}
 \includegraphics[width=.8\hsize]{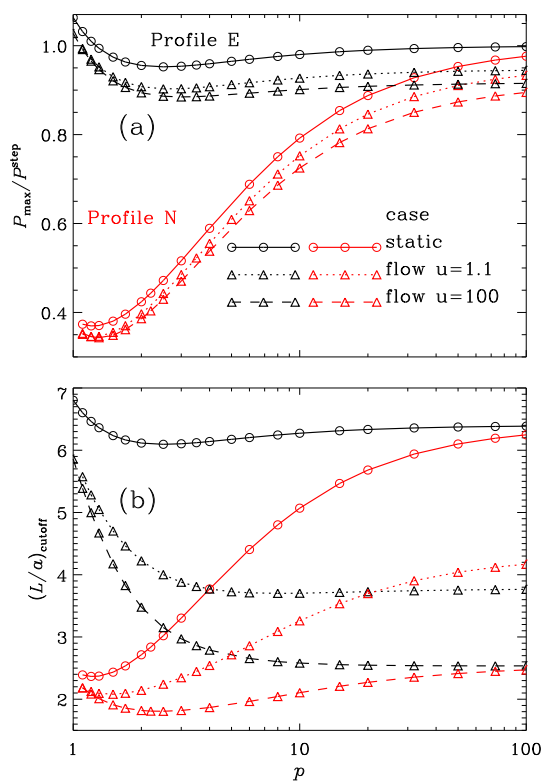}
 \caption{Dependence on density {  profile} steepness $p$ of (a) the maximum sausage period $P_{\max}$ and (b)
 the threshold length-to-radius ratio $\lacut$. The black (red) curves are for loops described by profile E (N).
 A density contrast of $25$ is adopted. The solid curves corresponds to the static case (flow magnitude $U_0 = 0$), while the dotted and dashed lines correspond to the {  cases with flow where the flow profile steepness $u$ is} $1.1$
 and $100$, respectively. In the cases {  with flow}, $U_0$ is fixed at $0.08~\vainf$. }
 \label{fig_period_p}
 \end{center}
\end{figure}

Figure~\ref{fig_period} presents the period of standing sausage modes, normalized by $a/\vainf$,
   as a function of the loop length in units of the loop radius $a$.
As in Fig.\ref{fig_DR}, profile N is adopted for the background density and flow speed.
A fixed density contrast $\rho_0/\rho_\infty$ of $25$ is chosen, while two values for the density {  profile} steepness,
   $p=2$ (the solid curves) and $p=100$ (dashed), are examined.
In addition to the static case (the black curves), a case {  with flow} (red) is also examined with the flow magnitude
   $U_0$ being $0.08 \vainf$ and flow {  profile} steepness $u$ being $100$.
Besides, the two dash-dotted straight lines represent $2L/v_{A0}$ and $2L/\vainf$, respectively.
In the static case, whichever $p$ the density profile adopts, the period curve
   follows the former asymptote in the thick-tube limit ($L/a \ll 1$),
   and terminates when intersecting the latter.
This happens because
   the phase speed $\vph$ tends to the internal Alfv\'en speed $v_{A0}$ when $ka \gg 1$,
   and attains the external one $\vainf$ as its maximum, beyond which sausage waves become leaky
   (see Fig.\ref{fig_DR}).
In the case {  with flow}, the red curves are not bounded by the two straight lines.
This arises because when $ka \gg 1$, the phase speed $\vph$ in the first (fourth) quadrant in Fig.\ref{fig_DR}
   tends to $v_{A0}+U_0$ ($-v_{A0}+U_0$).
At a given loop length, it can be seen that $P$ in the case {  with flow} is higher than its static counterpart.
However, while $P$ also increases with increasing $L$, the maximum period the trapped modes can reach, $\Pmax$, does
   not exceed the corresponding value in the static case,
   for the maximum allowed loop length is substantially shorter.
Despite the differences in the approaches adopted, the static computations (the black curves) agree closely with Fig.3 in NHM12 for the trapped modes.
In particular, the tendency for $P$ to be higher for steeper density profiles means
   that the step-function limit $\Pstep$ (Eq.(\ref{eq_Pmax_step})) may be the upper limit to the period
   that trapped modes may attain when profile N is adopted.
We note that for profile E, the same tendency for $P$ to increase monotonically with $L$ also holds, 
   for static and {  non-static} loops alike.
However, showing the relevant curves would make the graph too crowded and therefore we have omitted them.

How does the maximum period $\Pmax$ depend on the density profile steepness $p$?
Will it exceed $\Pstep$ if a different profile is chosen or in some particular range of $p$?
This is examined in Figure~\ref{fig_period_p}a where both profiles N (the red curves) and
    E (black) are plotted, and where $\Pmax/\Pstep$ instead of $\Pmax$
    is shown as a function of $p$.
Here the density contrast $\rho_0/\rhoinf$ is fixed at $25$.
In addition, the case with a fixed flow magnitude $U_0 = 0.08\vainf$
   is represented by the dotted and dashed curves, corresponding to
   a flow {  profile} steepness $u$ being $1.1$ and $100$, respectively.
For other choices of $u$, the results lie in between.
Several points are clear from Fig.\ref{fig_period_p}a.
First, for both profiles the effect of flow {  profile} steepness on $\Pmax$ is marginal: while $\Pmax$ slightly decreases with increasing $u$
   at a given $p$, $\Pmax$ in the case $u=100$ is smaller than $\Pmax$ for $u=1.1$ by no more than a few percent.
Moreover, $\Pmax$ in the case {  with flow} is always smaller
   than its static counterpart for all the examined density {  profile steepnesses},
   which is true also for both profiles.
Second, for {  profile E} the maximum period $\Pmax$ is insensitive to the density {  profile} steepness $p$.
In contrast, {  for profile N} $\Pmax$ shows a remarkably sensitive $p$-dependence.
Take the static case for instance, with $p$ ranging from $1$ to $100$,
   $\Pmax$ for profile E varies by $\lesssim 12\%$;
   however, $\Pmax/\Pstep$ for profile N increases considerably from $0.37$ at $p \sim 1$ to $0.98$ when $p=100$.
Third, one can see that for both profiles the dependence on $p$ of $\Pmax$ is not monotonic.
Take the static case for profile E for example, where with $p$ increasing from $1$, $\Pmax$ decreases rather than increasing
   first to a local minimum at $p \approx 2.5$ before rising again.
The same non-monotonic behavior takes place also for profile N, although the variation in $\Pmax$ at small $p$
   appears too shallow to distinguish.
The peculiar $p$-dependence is closely related to the $p$-dependence of $k_c^+$, given that $\Pmax$ is
   determined by $2\pi/(k_c^+ \vainf)$.

The trapping capabilities of loops are better shown by Fig.\ref{fig_period_p}b where
    the cutoff length-to-radius ratio $\lacut$ is plotted.
With the rest of the parameters chosen, this $\lacut$ is the maximum value a loop may have for it
    to support trapped standing sausage modes.
Consider static loops first.
One can see that for both profiles, the way in which $\lacut$ depends on the density {  profile} steepness $p$
    is identical to how the maximal period $\Pmax$ behaves, which is not surprising
    given that $\Pmax = 2\lacut (a/\vainf)$.
In a more intuitive way, the behavior of $\lacut$ indicates that for profile N, the general tendency is that
    the steeper the profile, the better a loop can trap sausage modes, in close agreement with NHM12.
However, for profile E, that a loop has a steeper density distribution does not necessarily mean it is a better
    \mbox{waveguide} for sausage modes.
Now consider {  non-static} loops, for which one can see that for both profiles, relative to the static case,
    introducing a flow with a magnitude being merely $0.08~\vainf$
    substantially reduces $\lacut$ at any given $p$.
This reduction is particularly prominent for large values of $p$.
Moreover, while the effect of the flow profile steepness $u$ is marginal as far as the maximum period is concerned,
    it is substantial in determining the maximum loop-to-radius ratio,
    with the tendency for both profiles being that $\lacut$ decreases with increasing $u$.
Take profile N (the red curves) for instance.
When $p=100$, $\lacut$ reads $6.25$ in the static case, and reads $4.17$ ($2.47$) when $u=1.1$ ($100$)
    in the non-static case.
The apparently unexpected behavior for steeper density or flow profiles to yield less strong
    trapping in certain cases derives from the intricate $p$- or $u$- dependence of
    the cutoff wavenumber $k_c^+$ and its combination with $\bar{k}_c^-$
    (see Fig.\ref{fig_DR}).
{\bf As illustrated by the red ($u=100$) and blue ($u=1.1$) curves in Fig.\ref{fig_stand_const},
    with increasing $u$ the $\omega-k$ curves in the first and second quadrants become increasingly asymmetric.
    While the $k_c^+$ values are similar in the two cases, the stronger asymmetry for $u=100$
    results in a $\bar{k}_c^-$ substantially larger.}
{\bf Consider} profile N and $p=100$ for example to examine the peculiar $u$-dependence.
    When $u=1.1$, one finds that $k_c^{+} = 0.526/a$ and $\bar{k}_c^{-} = 0.981/a$.
    However, when $u=100$, one finds that $k_c^{+}$ and $\bar{k}_c^{-}$ read 
    $0.549/a$ and $1.994/a$, respectively.

Figure~\ref{fig_period_p}a have a number of important implications for interpreting the quasi-periodic oscillations (QPOs)
    in flare lightcurves.
First, the period analytically derived for a piece-wise constant density profile (the step-function case,
    Eq.(\ref{eq_Pmax_step})) can be taken as an upper limit for the periods that
    standing sausage modes may attain.
This is true regardless of the particular density profile or whether a loop flow exists,
    meaning that these two complications do not broaden the period range
    of the QPOs that may be attributed to sausage modes.
Second, the seismological tool to probe the transverse density {  profile} steepness, proposed in NHM12 capitalizing on
    the sensitive dependence on the density {  profile} steepness of the maximum period,
    requires that the density profile be more properly described by profile N
    rather than profile E.
Given the importance of obtaining the information on the transverse fine structuring,
    we conclude that there is an imperative need to observationally distinguish between the two profiles.

\begin{figure}
\begin{center}
 \includegraphics[width=.8\hsize]{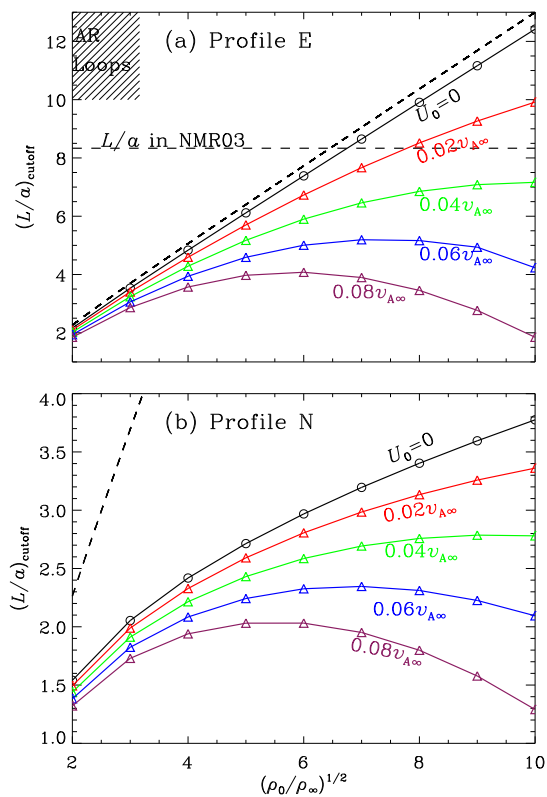}
 \caption{Dependence on density contrast $\rho_0/\rhoinf$ of threshold length-to-radius ratio $\lacut$ only below which can standing sausage modes be trapped. Both (a) profile E and (b) profile N are examined. The black solid curves are for the static case, while the colored curves represent the {  non-static} cases with
    the flow magnitude $U_0$ ranging from $0.02$ to $0.08~\vainf$. In the computations, both the density and flow {  profile} steepness, $p$ and $u$, are taken to be $2$. The dashed straight lines in both panels represent the threshold
 {  length-to-radius} ratio analytically derived in the static case for a step-function profile (Eq.(\ref{eq_la_cutstep})). Furthermore, the hatched area corresponds to the parameter range for typical active region (AR) loops. On the other hand, the horizontal dashed line represents the length-to-radius ratio of the flaring loop reported in~\citet{2003A&A...412L...7N}.}
 \label{fig_cutoff_ratio}
 \end{center}
\end{figure}

{  We can now further examine} the parameter range where trapped sausage modes are allowed.
This is done in Fig.\ref{fig_cutoff_ratio} which plots the maximum length-to-radius ratio $\lacut$
    as a function of density contrast $\rho_0/\rhoinf$, or more precisely the
    Alfv\'en speed ratio $\sqrt{\rho_0/\rhoinf}$,
    for both profiles E (upper panel) and N (lower),
    and for both static (black) and {  non-static} (colored) cases.
A series of values for the flow magnitude is presented, with $U_0$ ranging from
    $0.02$ to $0.08~\vainf$.
For illustrative purpose, the values for the density and flow {  profile} steepness, $p$ and $u$,
    are both chosen to be $2$.
In addition, the dashed lines in both panels represent $\lacut$ (Eq.(\ref{eq_la_cutstep}))
    attained in the step-function limit for static loops for comparison.
Each curve divides the $L/a - \rho_0/\rhoinf$ space into two regions,
    with trapped sausage modes permitted (prohibited) to its right (left).
While similar in format to Fig.2 in~\citet{2004ApJ...600..458A} where a density profile of step-function form
    is used, here the effects of both a smooth
    transverse profile and the presence of a loop flow are included.
One can see that with the chosen $p$,
    for static loops with profile E, the cutoff length-to-radius ratio (the black solid
    curve in Fig.\ref{fig_cutoff_ratio}a)
    does not differ significantly from the step-function limit (the dashed line),
    whereas for static loops with profile N, $\lacut$ is substantially smaller,
    with the deviation increasing with the density contrast (Fig.\ref{fig_cutoff_ratio}b).
Furthermore, introducing a loop flow leads to a reduction in $\lacut$ in general.
In particular, for both profiles E and N, when $U_0$
    exceeds $\sim 0.04~\vainf$, the dependence of $\lacut$ on $\rho_0/\rhoinf$
    is not monotonic but attains a maximum.
Let us examine what this means for flaring loops, for which we plot the horizontal dashed line
    in Fig.\ref{fig_cutoff_ratio}a corresponding to $L/a = 25~{\rm Mm}/3~{\rm Mm}$, taken from~\citet{2003A&A...412L...7N} (hereafter NMR03) and seen as being representative.
One can see that for profile E,
    a density contrast is required to exceed $\sim 45.6$ ($60.8$) for loops with
    $U_0$ being $0$ ($0.02~\vainf$)
    to host trapped sausage modes.
When $U_0 \gtrsim 0.04~\vainf$, however, the horizontal dashed line does not intersect
    the corresponding colored curves, meaning that if a flow with such a magnitude
    was present in the flaring loop analyzed in NMR03, the sausage mode would be in
    the leaky regime.
Note that this flow magnitude is not unrealistic: taking $\vainf$ to $\sim 3300$~\velunits\ derived
    for this particular event~\citep{2005A&A...439..727M},
    one finds that evaluating $0.04\vainf$ yields $\sim 130$\velunits.
As for profile N, even in the case of static loops,
    for them to trap standing sausage modes, it turns out that $\rho_0/\rhoinf$ has to exceed $\sim 2970$,
    which seems large even for flaring loops and is actually
    beyond the horizontal range used to plot Fig.\ref{fig_cutoff_ratio}.
Hence if profile N better describes the density profile of the flaring loop considered in~NMR03,
    then the standing mode identified therein should be a leaky one.

Figure~\ref{fig_cutoff_ratio} also allows us to say a few words on active region (AR) loops,
      for which the combination of typical length-to-radius ratios $L/a$
      and density contrasts $\rho_0/\rhoinf$ corresponds to the hatched area in Fig.\ref{fig_cutoff_ratio}a.
We note that $L/a$ is typically $\lesssim 0.05$
    (Table 1 in~\citeauthor{2002ApJ...576L.153O}~\citeyear{2002ApJ...576L.153O},
    also Figure 1 in~\citeauthor{2007ApJ...662L.119S}~\citeyear{2007ApJ...662L.119S})
    but taken here to be $\lesssim 0.1$ for safety.
The density contrast of AR loops relative to their ambient is difficult to yield,
    nevertheless, a range of 2 to $10$ is often quoted.
From Figs.\ref{fig_cutoff_ratio}a and \ref{fig_cutoff_ratio}b one expects that
    AR loops do not support trapped sausage modes: in the case of profile E,
    the hatched area is far from all the curves; in the case of profile N,
    it is beyond the range in which the curves are plotted.
Moreover, due to their mild density contrast, AR loops are unlikely to support observable leaky modes either.
This is because, as estimated by~\citeauthor{1975IGAFS..37....3Z}~(\citeyear{1975IGAFS..37....3Z},
    also Eq.(6) in~\citeauthor{2007AstL...33..706K}~\citeyear{2007AstL...33..706K})
    for static loops with a step-function density distribution,
    in the thin-tube limit the ratio of the damping time to wave period $\tau/P$ is approximately $(\rho_0/\rho_e)/\pi^2$,
    which evaluates to $\lesssim 1$ for AR loops.
Damped oscillations with such low quality factors would be difficult to detect.

\section{Summary}
\label{sec_conc}

The present work is motivated by two {  series} of studies in the context of coronal seismology.
First, while ubiquitous in coronal loops in general~\citep{2004psci.book.....A}
   and found in a number of oscillating loops in particular~\citep{2008A&A...482L...9O},
   field-aligned loop flows seem to have received insufficient attention as to their effects
   on the standing sausage modes.
Second, widely accepted to account for second-scale quasi-periodic oscillations (QPOs)
   in solar flare lightcurves, sausage modes have been shown recently by~\citet{2012ApJ...761..134N}
   to offer an additional diagnostic capability for inferring how steep the transverse
   density profile is.
The latter was made possible through the sensitive dependence of the sausage mode period on
   the density profile steepness.
We therefore are interested in assessing the combined effects of a field-aligned loop flow
   as well as smooth transverse density and flow profiles on the characteristics of standing sausage modes.
To this end, we work in the framework of zero-beta ideal {  magnetohydrodynamics} (MHD) and examine the linear
   sausage waves trapped in straight cylinders with field-aligned flow and enhanced density embedded in
   a uniform magnetic field.
Formulating the problem as a standard eigen-value one, we examine the dispersion diagrams
   and describe the procedure for constructing standing sausage modes.
Besides, we distinguish between two profiles, E and N, that both provide a smooth distribution connecting
   the values at loop axis (subscript $0$) and those at distances far from the loop (subscript $\infty$).
The end result is that, the maximum period $\Pmax$ that trapped {  standing} modes may attain,
   and the cutoff length-to-radius ratio $\lacut$ that a loop is allowed to reach for sausage modes to be trapped, depend {  on} the choice of the transverse profiles and a combination of dimensionless parameters
   $[\rho_0/\rhoinf, p; U_0/\vainf, u]$.
Here $p$ denotes the density {  profile} steepness, while $\rho_0/\rhoinf$ is the density contrast of the loop relative to
   its surroundings.
In addition, $U_0$ is the flow magnitude, $\vainf$ is the external Alfv\'en speed,
   and $u$ the flow {  profile} steepness.
For both profiles, the larger $p$ or $u$, the closer the profiles are to a step-function form.
{  Our main results are summarized as follows.}

For both profiles, the sausage mode period $P$ increases with increasing loop length. 
While at a given loop length $P$ in the {  non-static} case is higher than in the static case, 
    the maximum period $\Pmax$ never exceeds its static counterpart due to the cutoff length-to-radius
    ratio being substantially smaller.
{  For either profile, }with a flow magnitude $U_0 \lesssim 0.08~\vainf$, the maximum period $\Pmax$ is not substantially
    affected by the flow {  profile} steepness $u$, despite that $u$ varies significantly from $\sim 1$ to $100$.
The analytically expected sausage mode period for static loops with a step-function {  density} profile
    can be taken as an upper limit that sausage modes may attain in the trapped regime even for loops
    with smooth profiles and accompanied by field-aligned flows. 
Consequently, incorporating these two factors into sausage mode studies does not broaden the period range for the quasi-periodic oscillations in flare lightcurves that sausage modes can be responsible for.
The maximum sausage period $\Pmax$ depends sensitively on the density {  profile} steepness $p$ if profile N 
    is adopted to describe coronal loops, meaning that period measurements can be used to infer the
    information on fine structuring in this case. 
However, when profile E is adopted, the dependence of $\Pmax$ on $p$ is rather insensitive. 
In this sense, observationally distinguishing between the two profiles is crucial for establishing
   $\Pmax$ as a seismological tool for this purpose.

By far the most important effect a flow introduces is to reduce the capability of coronal loops for trapping sausage modes:
    it lowers the cutoff length-to-radius ratio $\lacut$ considerably relative to the static case. 
The {  flow profile} steepness $u$ is also important in this sense: $\lacut$ tends to decrease
    with increasing $u$.
Examining the parameter space subtended by length-to-radius ratio and density contrast, 
   Fig.\ref{fig_cutoff_ratio} shows that typical active region loops do not support trapped
   sausage modes, and are unlikely to support leaky modes with observable quality factors given
   their mild density contrast. 
As to the flaring loop reported by~\citet{2003A&A...412L...7N}, 
   the fundamental sausage mode identified therein is likely to be a leaky one if 
   profile N with $p=2$ (similar to a Gaussian) describes the transverse density distribution. 
This can also be said for profile E with $p=2$ when a flow with a magnitude stronger
   than $\sim 130$~\velunits\ was present in the loop.

\begin{acknowledgements}
    We thank the referee for his/her valuable comments, which helped improve this manuscript substantially.
    This research is supported by the 973 program 2012CB825601, National Natural Science Foundation of China (40904047, 41174154, 41274176, and 41274178), the Ministry of Education of China (20110131110058 and NCET-11-0305), and by the Provincial Natural Science Foundation of Shandong via Grant JQ201212.
\end{acknowledgements}


\begin{appendix}
\section{A Validation study of the BVPSUITE code}
\renewcommand{\theequation}{\thesection\arabic{equation}}

The eigen-value problem solver BVPSUITE seems new in the context of coronal seismology, 
   hence a study validating its accuracy seems
   in order. 
To this end, we carry out a series of computations in both slab and cylindrical geometries, and compare the numerically 
   derived dispersion curves with available analytic expectations. 
{\bf Given that these analytic expressions are derived in the limit of
   zero-beta MHD}, we focus on the same situation accordingly.

\begin{figure}
\begin{center}
 \includegraphics[width=.8\hsize]{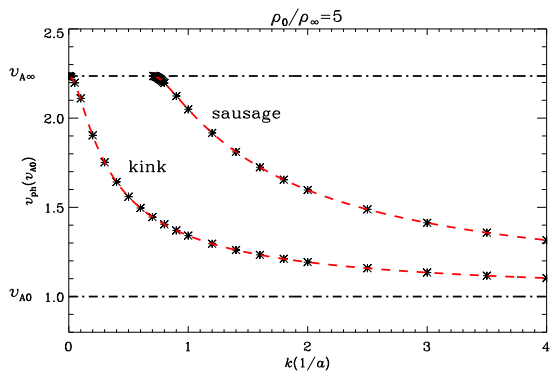}
 \caption{Dependence on the longitudinal wavenumber $k$ of the phase speed $\vph$ for a static cold slab
    with a symmetric Epstein density profile. 
 The asterisks represent the results obtained with BVPSUITE by numerically solving the eigen-value problem,
     while the red dashed lines represent the analytic results, Eqs.(\ref{eq_vali_slb_knk}) and (\ref{eq_vali_slb_ssg}).
 Here a density ratio $\rho_0/\rhoinf$ of $5$ is adopted.}
 \label{fig_vali_slab}
 \end{center}
\end{figure}

Let us start with the slab geometry {\bf and consider the static case. 
The magnetic slab} and the uniform equilibrium
   magnetic field $\bar{\vec{B}}$
   are both aligned with the $z$-axis. 
The equilibrium density $\bar{\rho}$ is structured in the $x$-direction.
Let us consider only the two-dimensional propagation in the $x-z$ plane. 
It then follows from the zero-beta MHD equations that the eigen-value problem for fast waves
   in the static case can be formulated as~\citep[e.g., Eq.(3) in][]{2005A&A...441..371T}
\begin{eqnarray}
 \vph^2\tilde{b}_x(x)=-\frac{v^2_{\rm A}(x)}{k^2}
    \left[\frac{\mathrm{d}^2}{\mathrm{d}x^2}-k^2\right]\tilde{b}_x(x) ,
    \label{eq_eigen_slab_bx} 
\end{eqnarray}
  together with the boundary conditions
\begin{eqnarray}
\tilde{b}_x(x\rightarrow \infty) \rightarrow 0,  
  \left\{ \begin{array}{lr}
                         \tilde{b}_x(x =0) =  0, & \mbox{sausage,} \\
			 \frac{\mathrm{d}\tilde{b}_x}{\mathrm{d}x}\left(x=0\right)=0, & \mbox{kink,}                     
                              \end{array}
  \right.
\end{eqnarray}
   where sausage and kink waves differ in their behavior at the slab axis ($x=0$).   
For the symmetric Epstein profile
\begin{eqnarray*}
   \bar{\rho}(x)=\rho_\infty+(\rho_0-\rho_\infty){\rm sech}^2\left(\displaystyle\frac{x}{a}\right)~, 
\end{eqnarray*}
   {\bf it is possible to solve the dispersion relation analytically for the phase speed $\vph$}~\citep{2003A&A...397..765C,2011A&A...526A..75M}.
For kink waves, it reads   
\begin{eqnarray}
  \left(\frac{\vph}{v_{\mathrm{A0}}}\right)^2 = 
    1+\frac{\sqrt{\zeta^{-2} + 4(ka)^2(1-\zeta^{-1})}-\zeta^{-1}}{2 (ka)^2} ,
  \label{eq_vali_slb_knk}  
\end{eqnarray}
   and for sausage waves, it reads
\begin{eqnarray}
  \left(\frac{\vph}{v_{\mathrm{A0}}}\right)^2 = 
    1+\frac{3\sqrt{9\zeta^{-2}-8\zeta^{-1}+4(ka)^2(1-\zeta^{-1})}-9\zeta^{-1}+4}{2 (ka)^2} ,
  \label{eq_vali_slb_ssg}  
\end{eqnarray}
   where $\zeta$ represents the density contrast $\rho_0/\rhoinf$.
Figure~\ref{fig_vali_slab} compares the dispersion diagram computed with BVPSUITE (the asterisks) with the analytic expressions (the red dashed curves) for a representative density contrast $\rho_0/\rhoinf$ being $5$. The internal and external Alfv\'en speed, $v_{\mathrm{A0}}$ and $\vainf$, are represented by the two horizontal dash-dotted lines. It can be seen that the analytic results are exactly reproduced. In fact, we have carried out a series of comparisons covering an extensive range of density ratios,
   and found exact agreement without exception.

{\bf Now move on to the cylindrical case, but still restrict ourselves to the static case for the moment.
We solve} the same eigen-value problem in the text, Eqs.(\ref{eq_eigen_br}) 
  and (\ref{eq_eigen_BC}), but now take the background flow $\bar{U}$ to be zero.
In addition, we take $p$ to be infinity, corresponding to a discontinuous distribution of the equilibrium
   density.
To facilitate the validation study, we focus on sausage waves given the availability of their analytic behavior both in the  
   neighborhood of the cutoff wavenumber and for large wavenumbers.
In the former, the dispersion behavior in terms of angular frequency $\omega$ can
   be expressed as \citep[Eq.(53) in][]{2014ApJ...781...92V}
\begin{eqnarray}
 \frac{\Delta\omega}{\Delta k}
   = \vainf 
   \frac{1-\ln\left(\frac{k_c^2 a^2}{2}\left|\frac{\Delta\omega}{k_c \vainf}-\frac{\Delta k}{k_c}\right|\right)}{{\rho_0}/{\rhoinf}-\ln\left(\frac{k_c^2 a^2}{2}\left|\frac{\Delta\omega}{k_c \vainf}-\frac{\Delta k}{k_c}\right|\right)} ,
   \label{eq_vali_cyl_arndkc}
\end{eqnarray}
   where $\Delta\omega = \omega - k_c \vainf$, $\Delta k = k-k_c$,
   and $k_c$ is the cutoff wavenumber given by Eq.(\ref{eq_kc_cut_step}).
For Eq.(\ref{eq_vali_cyl_arndkc}) to be valid, one nominally requires that $\Delta k/k_c \ll 1$.
On the other hand, when $ka \gg 1$, the dispersion relation Eq.(8b) in~\citet{1983SoPh...88..179E}
   can be shown to yield
\begin{eqnarray}
 \vph \approx v_{\mathrm{A0}}\sqrt{1+\frac{j_{1,l}^2}{k^2 a^2}} ,
 \label{eq_vali_cyl_bigk}
\end{eqnarray}
   where $j_{1,l}$ ($l=1, 2, \cdots$) denotes the $l$-th zero of $J_1$
   and $l$ denotes the infinite number of sausage branches.
For the first branch which we compute, $j_{1,1} = 3.83$.   

\begin{figure}
\begin{center}
 \includegraphics[width=.8\hsize]{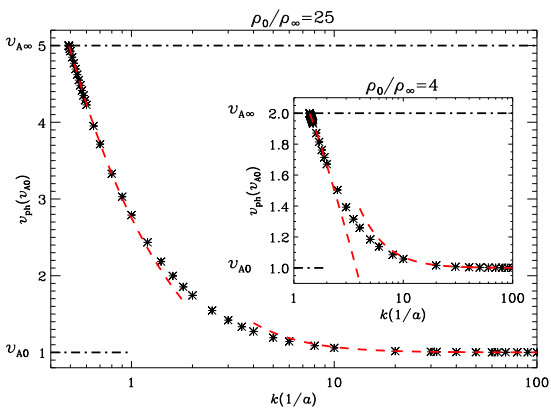}
 \caption{Similar to Fig.\ref{fig_vali_slab} but for sausage waves supported by static cylinders with discontinuous density distribution.
 The red dashed curves represent the analytic results in the vicinity of the cutoff wavenumber
 (Eq.\ref{eq_vali_cyl_arndkc}) and for big wavenumbers (Eq.\ref{eq_vali_cyl_bigk}). Here two density ratios, $4$ and $25$, are adopted.
 }
 \label{fig_vali_cyl}
 \end{center}
\end{figure}

Figure~\ref{fig_vali_cyl} presents the dispersion curves expressing the phase speed $\vph$ as a function of longitudinal wavenumber $k$.
For illustrative purposes, we present the results for two density contrasts, one large ($\rho_0/\rhoinf = 25$)
   and the other rather mild ($\rho_0/\rhoinf = 4$, inset).
The results computed with BVPSUITE are given by the asterisks,
   and the analytic results are represented by the red dashed curves.
The two horizontal dash-dotted lines represent the internal and external Alfv\'en speeds.   
Evidently, the numerical results excellently capture the cutoff wavenumber,
    and agree remarkably well with the analytic results for the appropriate wavenumber ranges.
Actually, this can be said for all the tests we performed, which cover an extensive range of density ratios.

{\bf Our next validation study pertains to sausage waves supported by cold cylinders with flow.
To this end we start with the comprehensive study by~\citet{1992SoPh..138..233G} where the sophisticated equilibrium configuration
    takes account of a background flow, and azimuthal components of the equilibrium velocity and magnetic field.
If neglecting these azimuthal components and specializing to a piece-wise constant distribution for both the equilibrium density
    and flow speed, one finds that Eq.(18) in~\citet{1992SoPh..138..233G} simplifies to 
\begin{eqnarray}
 \frac{1}{r}\frac{\mathrm{d}}{\mathrm{d}r}\left(r \frac{\mathrm{d} \tilde{P}}{\mathrm{d}r}\right) - m^2 \tilde{P} = 0 ,
 \label{eq_vali_G18}
\end{eqnarray}
    where $\tilde{P}$ denotes the Fourier amplitude of the total pressure perturbation.
Equation~(\ref{eq_vali_G18}) is valid both inside and outside the cylinder, and $m^2$ is defined as
\begin{eqnarray*}
   m_0^2 = \frac{k^2 \vain^2 -(\omega - k U_0)^2}{\vain^2}, \skip 0.2cm
   m_\infty^2 = \frac{k^2 \vainf^2 -\omega^2}{\vainf^2}, 
\end{eqnarray*}
   in which we have assumed that the ambient corona is static.
For the simple configuration in question, Eq.(\ref{eq_vali_G18}) is analytically solvable in terms of Bessel function $J_0$ ($K_0$) 
    inside (outside) the cylinder for trapped modes.
A dispersion relation then follows from the continuity of the transverse Lagrangian displacement 
    and total pressure perturbation at the cylinder boundary~\citep[see also, e.g.,][]{2003SoPh..217..199T},
\begin{eqnarray}
 \rhoinf \left(\vainf^2 - \vph^2\right) n_0 \frac{J_0'(n_0 a)}{J_0(n_0 a)}
     = \rho_0 \left[\vain^2 -(\vph - U_0)^2\right] m_\infty \frac{K_0'(m_\infty a)}{K_0(m_\infty a)} ,
 \label{eq_vali_cyl_flow_disp}     
\end{eqnarray}
   where $n_0^2 = -m_0^2$.
Besides, the prime denotes the derivative of Bessel function with respect to its argument, e.g.,
   $K_0'(m_\infty a) = \mathrm{d} K_0(\eta)/\mathrm{d}\eta$ with $\eta = m_\infty a$.

\begin{figure}
\begin{center}
 \includegraphics[width=.8\hsize]{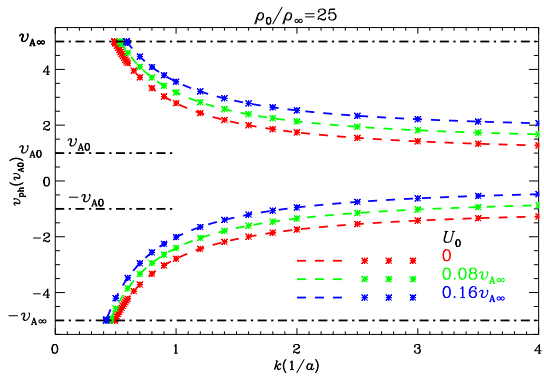}
 \caption{Dependence on the longitudinal wavenumber $k$ of the phase speed $\vph$ for a cold cylinder with flow where
    the transverse distributions of both the equilibrium density and field-aligned flow adopt
    a step-function form.
 The asterisks represent the results obtained with BVPSUITE by numerically solving the eigen-value problem,
     while the dashed lines represent the solution to the analytically derived dispersion relation (Eq.(\ref{eq_vali_cyl_flow_disp})).
 The horizontal dash-dotted lines correspond to the internal and external Alfv\'en speeds.
 Here three different values of the internal flow speed $U_0$ are examined for a density ratio $\rho_0/\rhoinf$ being $25$.
 }
 \label{fig_vali_cyl_flow}
 \end{center}
\end{figure}

Figure~\ref{fig_vali_cyl_flow} presents the dependence of the phase speed $\vph$ on the axial wavenumber $k$
    for a representative density contrast $\rho_0/\rhoinf$ being $25$.
For illustrative purpose, we examine three values of the internal flow speed $U_0$, 
    namely, $0$ (red), $0.08$ (green), and $0.16$ (blue) times the external Alfv\'en speed $\vainf$.
The horizontal dash-dotted lines represent the internal and external Alfv\'en speeds.
The asterisks give the results from solving Eqs.(\ref{eq_eigen_br}) and (\ref{eq_eigen_BC}) in the text with BVPSUITE,
    where both the density and flow speed profile steepnesses, $p$ and $u$, are taken to be infinity.
For comparison, the dashed curves represent the solutions to the algebraic dispersion relation, Eq.(\ref{eq_vali_cyl_flow_disp}).
One can see that the two sets of solutions agree with each other remarkably well.
As a matter of fact, the agreement is found for all the tests we conducted, where we examined an extensive range of
    density contrasts and flow magnitudes.
}

In closing, let us mention that at this stage of its development, the Matlab eigen-value problem solver
   BVPSUITE cannot find complex eigen-values, thereby limiting its use to trapped modes. 
Despite this, given that BVPSUITE is publicly available, easy to use with its friendly graphical user interface, 
   this accurate code may find a wider application to problems one encounters 
   in coronal seismology. 

\end{appendix}

\end{document}